\documentclass[12pt]{revtex4}
\usepackage{graphicx}
\usepackage{times}

\begin{document}

\title {ELECTRON SCATTERING FROM GASEOUS OCS($^1\Sigma$): COMPARING
COMPUTED ANGULAR DISTRIBUTIONS AND ELASTIC CROSS SECTIONS WITH
EXPERIMENTS.}
\author{F.A. Gianturco\footnote{Corresponding author; e.mail address:
fa.gianturco@caspur.it. Fax: +39-06-4991.3306.}}
 \affiliation{Department of Chemistry, The University of Rome La Sapienza and CNISM,
 Piazzale A. Moro 5, 00185 Rome, Italy.}
\author{T. Stoecklin}
\affiliation{Laboratoire de Physico-Chimie Moleculaire, 351, Cours
de la Lib\'{e}ration, F-33405 Talence, France.}

\begin{abstract}
Differential cross sections are computed for the title polar
molecule using  static interaction,  exchange forces and
correlation-polarisation effects as described in detail in the main
text. The dipole effect is also reported via the dipole Born
correction procedure and the final angular distributions are
compared with existing experimental data. The shape and location of
the prominent low-energy resonance are  computed and compared with
experiments. The comparison shows that the present treatment of the
interaction forces and of the quantum dynamics can indeed afford
good agreement between measured and computed quantities for a
multielectron target as OCS.
\end{abstract}

\maketitle

\newpage

\section{Introduction}
The carbon oxysulfide (OCS) molecule is well known to be of
considerable importance for the role it plays in the carbon
chemistry and sulphur chemistry chains of reactions in molecular
astrophysics environments like those in the diffuse and dark
interstellar clouds (DISC)\cite{1}. It is also of interest in the
realm of cold molecular plasmas because of its function as a
reaction quencher into the plasma formation. It follows, therefore,
that collisional processes induced by low-energy electrons in this
molecular gas have triggered the interest of both experimentalists
and theoreticians, intrigued by the possible role that the permanent
dipole of this target molecule can play in characterizing the
scattering attributes.

On the theoretical side, earlier calculations were carried out by
Lynch et al.\cite{2} using their continuum scattering model, with
which they found the presence of a $\Pi$ symmetry shape resonance
around 1.15 eV and a sharp increase at threshold of the $\Sigma$
symmetry partial cross section. The differential cross sections
(DCSs) for elastic scattering and for vibrational excitations were
measured  by Sohn and coworkers \cite{3}, where the authors also
observed a $\pi ^{\ast}$ resonance and a strong excitation of the
bending mode associated to dipole coupling and to a possible bound
state of $(OCS^-)^{\ast}$. Further measurements are due to
Szmytkowski et al. \cite{4} who measured the total, integral cross
sections of the title molecule and confirmed a $\Pi$-type resonance
around 1.15 eV. More recent measurements have been carried out by
the Japanese group \cite{5}, who compared similar measurements with
electrons and positrons as projectiles and confirmed the presence of
the $\pi ^{\ast}$ resonance around 1.15 eV. The vibrational
excitation of OCS has been measured by another study from the same
group \cite{6}. Calculations of elastic cross sections, integral and
differential, and of the momentum transfer cross sections have been
performed by Bettega et al. \cite{7}, who employed the Schwinger
multichannel method with pseudopotentials within the static-exchange
approximation and examined a range of energies from 5 to 50 eV.

A combined theoretical-experimental study has been conducted by
Michelin et al. \cite{8} who carried out both measurements and
calculations in the energy regime from 0.4 eV to 600 eV. Further
calculations of both integral and differential cross sections were
reported very recently by Bettega et al. \cite{9}, who employed
again the Schwinger multichannel method with pseudopotentials,adding
the polarization interaction to their static-exchange initial
scheme. A further "addendum" to the above results was recently
published with regard to the $\Sigma$-symmetry partial cross
sections \cite{10}.

In the present study we intend to discuss in some detail the
behavior of the computed angular distributions with respect to the
available experiments over a very broad range of collision energies,
using an entirely different method as that employed by the earlier,
just as extensive, calculations of references \cite{8,9,10}. The
following Section describes briefly our theoretical and
computational approach while Section III reports our results for
both the partial integral cross sections and the differential cross
sections. The last Section IV summarizes our present conclusions.

\section{The Theoretical Modelling}
\subsection{The separable exchange representation}
One of the major requests from accurate calculations of low energy
electron-molecule scattering observables is the correct description
of the non-local part of the interaction potential.  For
electronically elastic collisions, a physically realistic treatment
usually starts with the static interaction and views the exchange
potential as a correction to the latter due to the imposition of
anti-symmetry on the total electronic wavefunction that describes
the  scattered electron plus the target bound electrons.  Even in
this case, however, the correct treatment of non-local forces for
polyatomic molecules still represents a rather formidable task and
therefore increasing use is being made of separate representations
which expand the exchange kernel in terms of products of
one-electron functions \cite{11}.   This approach will be adopted in
the present paper, following our extension of this method to
polyatomic molecular targets \cite{12,13}. It has the advantage of
being a non-empirical method for treating exchange forces while
still offering computational savings with respect to the exact
solution of the integro-differential equations of non-local
electron-molecular scattering theory \cite{14}.  The earlier work on
separate exchange in electron collision with small diatomics  has
been rather encouraging since fairly small exchange basis sets added
little more computer time than a local potential calculation
\cite{11,12} and therefore we use such an approach on the linear
polyatomic target of the present study.

One starts by approximating the exchange potential $W(\mathbf{r},
\mathbf{r}_e)$ by the truncated separate form

\begin{equation}
W(\mathbf{r},\mathbf{r}_e)\approx
\sum_{\alpha,\beta}^{N}\chi_{\alpha}(\mathbf{r})\mathbf{W}_{\alpha,\beta}\chi_{\beta}(\mathbf{r}_e)
\end{equation}

\noindent where the $\{\chi_{\alpha}\}$ now constitute an
additional, new set of Cartesian GTOs not orthogonal to each other,
nor to the occupied molecular orbitals (MOs) of the target SCF basis
set which describes the ground state of the target molecule. The
vector coordinates $(\mathbf{r})$ and $(\mathbf{r}_e)$ describe the
positions of the bound and of the scattered electrons, respectively.
The $W_{\alpha,\beta}$ now represents the two-electron interaction
operator over the truncated, discrete basis.

When performing the separable expansion (1) care should be taken of
how to specify  the exchange basis functions $\{\chi_{\alpha}\}$ in
order to avoid taking too  large an expansion so that the
calculations become too massive or begin to suffer from linear
dependency effects. We have carefully checked this aspect of the
problem by analysing in each case the corresponding eigenvectors of
the overlap matrix and modifying the basis set accordingly.

The required exchange matrix elements for the bound MOs of the
target, taken to be a closed-shell structure, are therefore given by
first calculating the following matrix elements ,via a standard
bielectronic subroutine

\begin{equation}
\tilde{\mathbf{B}}_{\gamma \tau} = \int d\mathbf{r}\int
d\mathbf{r}_e\varphi
_{\gamma}(\mathbf{r})W(\mathbf{r},\mathbf{r}_e)\varphi_{\tau}(\mathbf{r}_e)
\end{equation}

In the above equation $\{\varphi _{\gamma}\}$ denotes the set of
doubly occupied self-consistent-field (SCF) target MOs. Hence, using
(1) one can further write

\begin{equation}
\tilde{\mathbf{B}}_{\gamma \tau} = \sum_{\alpha,\beta}^{N}\int
d\mathbf{r}\varphi
_{\gamma}(\mathbf{r})\chi_{\alpha}(\mathbf{r})W_{\alpha,\beta} \int
d\mathbf{r}_e\varphi _{\tau}(\mathbf{r}_e)\chi_{\beta}(\mathbf{r}_e)
\end{equation}

\noindent our further compact the overlap integrals

\begin{equation}
\tilde{\mathbf{B}}_{\gamma \tau} = \sum_{\alpha,\beta}^{N}S_{\gamma
\alpha}W_{\alpha,\beta} S_{\beta \tau}
\end{equation}

One can finally obtain the bound-continuum scattering matrix from
the separable representation by writing it down via the following
product of the matrices already defined in the above equations:

\begin{equation}
\mathbf{W = S}^{-1}\mathbf{\tilde{K}S}^{-1}
\end{equation}

\noindent where the $S_{\gamma \alpha}$ are the overlap matrix
elements between the additional GTO set and the original expansions
describing the bound MOs.

\subsection{The scattering equations}
Within a single-centre expansion (SCE) of the continuum wavefunction
and of the interaction potential, the use of the present
static-separable-exchange  approximation gives rise to a set of
coupled integro-differential equations

\begin{equation}
\left\{\frac{d^2}{dr_e^2}+ \frac{l(l+1)}{r_e^2}\right\}
u_{ll_0}(r_e)=\sum_{l'} \left\{U_{ll'}(r_e)u_{l'l_0}(r_e)+\sum
_{\alpha \beta}\Phi_{\alpha}^l(r_e)W_{\alpha \beta}\int
dr\Phi_{\beta}^l (r)u_{l'l_0}(r)\right\}
\end{equation}

\noindent where

\begin{equation}
U_{ll'}(\mathbf{r}_e)=\int
S_l^m(\hat{r}_e)V(\mathbf{r}_e)S_{l'}^{m'}(\hat{r}_e)d\hat{r}_e
\end{equation}

\noindent and

\begin{equation}
\Phi_{\alpha}^{\ell} (r_e)=(r_e) \int d\hat{r}_3\varphi_{\alpha}
(\mathbf{r}_e)S_l^m(\hat{r}_e)
\end{equation}

\noindent which integrates over real spherical harmonics to yield
the radial part of each new GTO function. Here the $\varphi$'s are
the orbital used in equation (2).

Furthermore

\begin{equation}
S_l^{m,p}(\hat{r}_e)=\frac{i}{\sqrt{2}}\{Y_l^m(\hat{r}_e)\pm
(-1)^pY_l^{-m}(\hat{r}_e)
\end{equation}

\noindent with the parity index p=0 or 1 and where the same equation
(8) also holds for the $r$ variable in (6).

One can now express the solution as a linear combination of
homogeneous and inhomogeneous terms:

\begin{equation}
u_{ll_0}(r_e)=u_{ll_0}^0(r_e)+\sum
_{\alpha}u_l^{\alpha}(r_e)C_{l_0}^{\alpha}
\end{equation}

\noindent where

\begin{equation}
(k^2-H_0^{l}) u_{ll_0}^0(r_e)=\sum _{l'}U_{ll'}(r_e)u_{l'l_0}(r_e)
\end{equation}

\begin{equation}
(k^2-H_0^{l}) u_{l}^{\alpha}(r_e)=\sum
_{l'}U_{ll'}(r_e)u_{l'}^{\alpha}(r_e)+\Phi_{\alpha}^l (r_e)
\end{equation}

The coefficients $C_l^{\alpha}$ are found to satisfy a set of linear
equations

\begin{equation}
\sum_{\beta}A_{\alpha\beta}C_{l_0}^{\beta}=B_{\alpha l_0}
\end{equation}

\noindent where

\begin{equation}
A_{\alpha\beta}=\delta_{\alpha\beta}-\sum_{l'\gamma}W_{\alpha
\gamma}\int \Phi_{\beta}^{l'}(r_e)u_{l'}^{\beta}(r_e)dr_e
\end{equation}

\noindent and

\begin{equation}
B_{\alpha l_0}=\sum_{l'\beta}W_{\alpha \beta}\int _0^0
\Phi_{\beta}^{l'}(r_e)u_{l'l_0}^0(r_e)dr_e
\end{equation}

The final numerical integration of the ensuing Volterra equations
was then carried out  as already described in Jain and Norcross
\cite{15}.

\section{Computed and Measured Cross Sections}

\subsection{Computational details}
The interatomic distances were fixed at their experimental values of
$R_{oc}= 2.196\,\, a_0$ and $R_{cs}= 2.941 a_0$. The two components
of the dipole polarisability tensor were taken also, in the
asymptotic interaction, to be the experimental values: $\alpha_0
(R_{eq}) =35.1 a_0^3$ and $\alpha_2 (R_{eq}) =17.5 a_0^3$. The above
choice was dictated by our desire to realistically treat low-energy
effects which are known to be strongly affected by polarisabilities.

To describe the target electrons in the neutral ground state we
employed a double-zeta D95V basis set plus polarization orbitals as
in the expansion described as D95V$^{\ast}$ \cite{16}. The obtained
total energy was therefore -1020.4122528 hartrees, with a dipole of
0.32 a.u. and a quadrupole of 0.87 a.u., to be computed with the
experimental values of 0.27 a.u. for the dipole \cite{4}. No
experimental value of the molecular quadrupole was found to be
available.

The radial grid chosen for the Volterra integrals went from 0.005
$a_0$ to 10.0 $a_0$, beyond which the integration was extended out
to 1,000 $a_0$ using the asymptotic static multipoles plus the
polarization potential. The multipolar went up to $\lambda
_{max}$=70 while the bound orbitals were expanded around the center
of mass up to $l_{max}$=118.

Table I reports the additional GTO's employed to describe the
separable exchange and used in the dynamical calculations outlined
in the previous Section.

\begin{table*}
\begin{center}
 \caption{\small{Basis set functions (types, locations and exponents) for the continuum
orbitals.}}
\begin{tabular}{c|c|c|c}
\hline
\multicolumn{1}{l}{location: on the O} &\multicolumn{1}{c}{on the C} & \multicolumn{1}{c}{on the S}& \multicolumn{1}{c}{on the Center of mass}\\
\hline
\multicolumn{1}{c}{$\sum$:} &\multicolumn{1}{c}{$\sum$:} & \multicolumn{1}{c}{$\sum$:}& \multicolumn{1}{c}{$\sum$:}\\
 \textbf{s:}8.0,4.0,2.0,& \textbf{s:}8.0,4.0,2.0,&
 \textbf{s:}8.0,4.0,2.0,& \\
1.0,0.5,0.25&1.0,0.5,0.25,0.12&1.0,0.5,0.25,0.12&\\
\textbf{p}$_\mathbf{z}$:2.0,1.0,0.5,0.25&\textbf{p}$_\mathbf{z}$:2.0,1.0,0.5,0.25
&\textbf{p}$_\mathbf{z}$:2.0,1.0,0.5,0.25 & \\
\textbf{d}$_\mathbf{xz}$:1.0,0.5,0.25&\textbf{d}$_\mathbf{xz}$:1.0,0.5,0.25,0.12
& \textbf{d}$_\mathbf{xz}$:1.0,0.5,0.25,0.12 & \\
$\Pi$:&$\Pi$: & $\Pi$:& $\Pi$:\\
\textbf{p}$_\mathbf{x}$:32.0,16.0,8.0, &&
\textbf{p}$_\mathbf{x}$:32.0,16.0,8.0, &
\textbf{p}$_\mathbf{x}$:32.0,16.0,8.0,4.0, \\
4.0,2.0,1.0,0.5 &&4.0,2.0,1.0,0.5 &2.0,1.0,0.5\\
 \textbf{d}$_\mathbf{xz}$:4.0,2.0,1.0,0.5,0.25
&& \textbf{p}$_\mathbf{xz}$:4.0,2.0,1.0,0.5,0.25 &
\textbf{d}$_\mathbf{xz}$:4.0,2.0,1.0,0.5,0.25 \\
$\Delta$:& $\Delta$: & $\Delta$:& $\Delta$:\\
\textbf{d}$_\mathbf{xx}$:16.0,8.0,4.0,2.0, & &
\textbf{d}$_\mathbf{xx}$:16.0,8.0,4.0,2.0, &\textbf{d}$_\mathbf{xx}$:16.0,8.0,4.0,\\
1.0,0.5,0.25 & &1.0,0.5,0.25 &4.0,2.0,1.0,0.5\\
 \textbf{f}$_\mathbf{x^2z}$:4.0,2.0,1.0,0.5,0.25 &
&\textbf{f}$_\mathbf{x^2z}$:4.0,2.0,1.0,0.5,0.25 & \textbf{d}$_\mathbf{xz}$: 4.0,2.0,1.0,0.5,0.25\\
 \hline
\end{tabular}
\end{center}
\end{table*}

The single-center partialwave expansion of the discretized,
continuum orbitals employed to treat exchange involved $l_{max}$=15
for the $\sum$ symmetry, $l_{max}$=20 for the $\Pi$ symmetry and
$l_{max}$=15 for the $\Delta$ symmetry.

The above set of functions was the result of a series of numerical
tests at a set of chosen energies where we increased the number and
quality of the additional GTO's until the corresponding, final
K-matrix elements varied by less than 0.1\%.

\subsection{The integral cross section}
The results reported by figure 1 analyze the partial and total
integral, elastic cross sections (rotationally summed) for their
overall energy dependence over the broad range of energies which is
experimentally available \cite{3,4,5}, but we only report a
comparison with some of the data in order not to crowd the figure
excessively. We also show in the lower panel the energy behavior of
the individual partial cross sections for the contributing partial
symmetries.

\begin{figure}
\begin{center}
\includegraphics[scale=0.50]{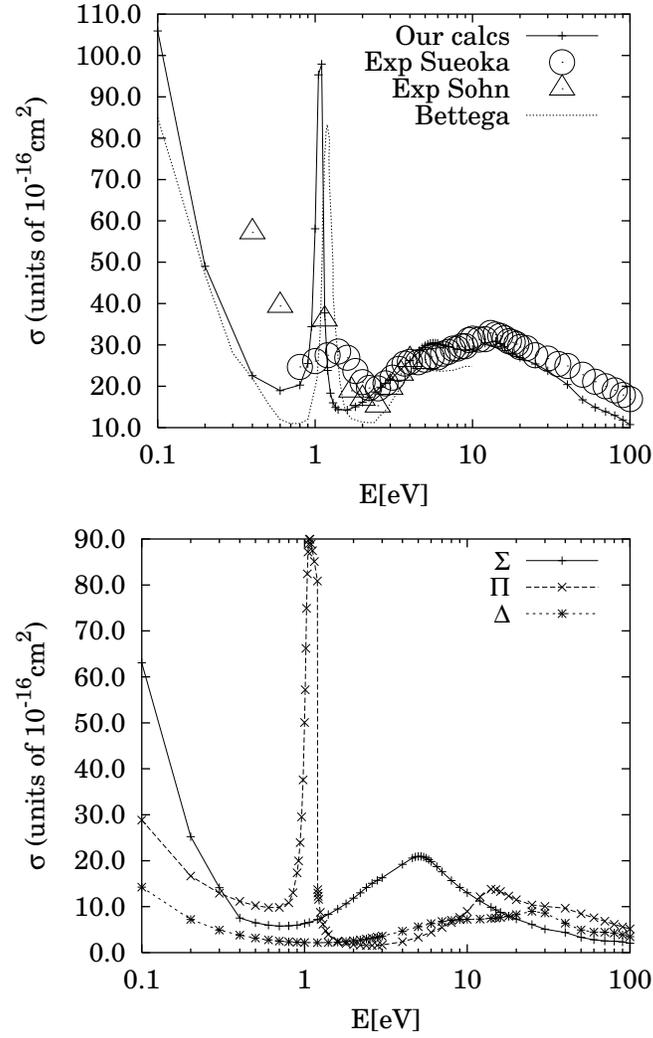}
\caption{Upper panel: Computed and measured total elastic integral
cross sections as a function of energy. Solid line: present results:
Dots: calculations from ref.\cite{9}. Open circles: expt.s from
ref.\cite{3}; open triangles: expt.s from ref.\cite{5}. Lower panel:
partial cross sections for the $\Sigma$-(solid line), $\Pi$-(dashes)
and $\Delta$-(dots) symmetries.}\label{figure1}
\end{center}
\end{figure}

It is reassuring to see that the present calculations follow closely
the experimental findings from below 1.0 eV of energy and up to 100
eV: this is a rather good result considering the complexity of the
molecule and the broad range of energies spanned by the
calculations. Furthermore, we see that the resonance position is
obtained reasonably well from our calculations, albeit with a
sharper peak due to the absence of vibrational averaging, which also
come very close to the recent calculations of Bettega et al.
\cite{9} that employed an entirely different method for obtaining
it.

The lower panel in the figure reports the energy dependence of the
various partial cross sections which contribute to the scattering
process. The calculations clearly show that:
\begin{enumerate}
  \item the resonance around 1 eV is due to the $\Pi$ symmetry i.e. to the expected
  dynamical (centrifugal) trapping of the electron into a
  $\pi^{\ast}$ metastable orbital by the $l$=1 angular momentum barrier;
  \item there is a second $\pi^{\ast}$ resonance at higher energies
  ($\sim$10 eV) which is a broader one and which is also seen in the
  experimental data;
  \item the $\Sigma$-symmetry exhibits the expected dominance of
  s-wave scattering as one moves down to low energies \cite{10} and
  the corresponding cross section goes through a Ramsauer-Townsend
  minimum around 0.7 eV, as also analyzed and discussed by recent calculations of Bettega et al.
  \cite{10};
  \item the $\Sigma$-symmetry cross section also suggests the
  presence of a $\sigma^{\ast}$ resonance around 4-5 eV which is
  also seen in the experiments (upper panel) as a shoulder on the
  raising cross section energy dependence around 4 eV.
In Fig. 2 we report the behaviour of the eigenphase sum associated
with the $\Sigma$ symmetry component. We see that, at low energies,
this quantity goes through zero, as in ref.\cite{10}, and therefore
it suggests the presence of an RT minimum in the elastic cross
section. It further increases from zero energy values and goes
through a maximum as shown by similar data in ref.\cite{10}.

In conclusions, the present calculations for partial and total
integral cross sections indicate that our present approach can
reliably describe the experimental behavior over a very broad range
of energies and can also reproduce earlier calculations very closely
to their findings as far as partial symmetry contributions are
concerned.

The previous literature \cite{8,9,10} also discusses the possibility
of having a virtual state close to zero energies detected by a
negative value of the scattering length. Since the present system is
a polar target, no modified effective range theory can be applied
\cite{16} and  no scattering length could therefore be uniquely
defined. Different procedures therefore need to be used to extract
information on the possible existence of virtual states \cite{18} in
polar targets. In the present case, however, we decided that such a
study was outside the scope of our current aims.

\subsection{Angular distributions at low energies}
We have shown above that the well marked, low energy resonance
appears, but in experiments and calculations, around 1 eV of
collision energy, and therefore it becomes of interest to also see
what the angular distributions look like across that range of
energies. The results of our calculations, and their comparisons
with the existing experiments, are shown in the various panels of
figure 3 and figure 4 where the energies vary from 0.4 eV up to 3.0
eV. The experiments are those of Sohn et al. \cite{3} and from
Tanaka  \cite{5}, respectively marked by open circles and open
triangles in all the panels, while the calculations (dashed curves)
are from reference \cite{8}. The following comments could be made:
\begin{enumerate}
  \item at the lowest collision energies the scattering is dominated
  by the weak dipole interaction which causes a marked forward
  scattering behavior, which is also well reproduced by our
  calculations. On the other hand, in the larger angle region the
  agreement becomes less satisfactory, possibly due to our model
  treatment of short-range correlation effects. We also see that at
  1.15 eV the earlier calculations \cite{8} also do not agree well with
  experiments.
  \item as the collision energy moves across the position of the
  strong $\pi^{\ast}$ resonance, as shown by the panels of figures 3
  and 4, we see that the angular distributions become much flatter
  from $\vartheta \sim 20^{\circ}$ and out to 140$^{\circ}$, with a
  markedly reduced forward peak, with the latter resuming as soon as
  the energy leaves the resonance region (e.g. see  panels in
  figures 3 and 4).
  \item We also see from Fig. 4 that, while the agreement with
experiments is good at 2.0 eV, and remains reasonableat 2.5 - 3 eV,
indicating numerical convergence of our present results to the same
level of reliability with respect to the experiments.
\end{enumerate}

\begin{figure}
\begin{center}
\includegraphics[scale=0.50,angle=-90]{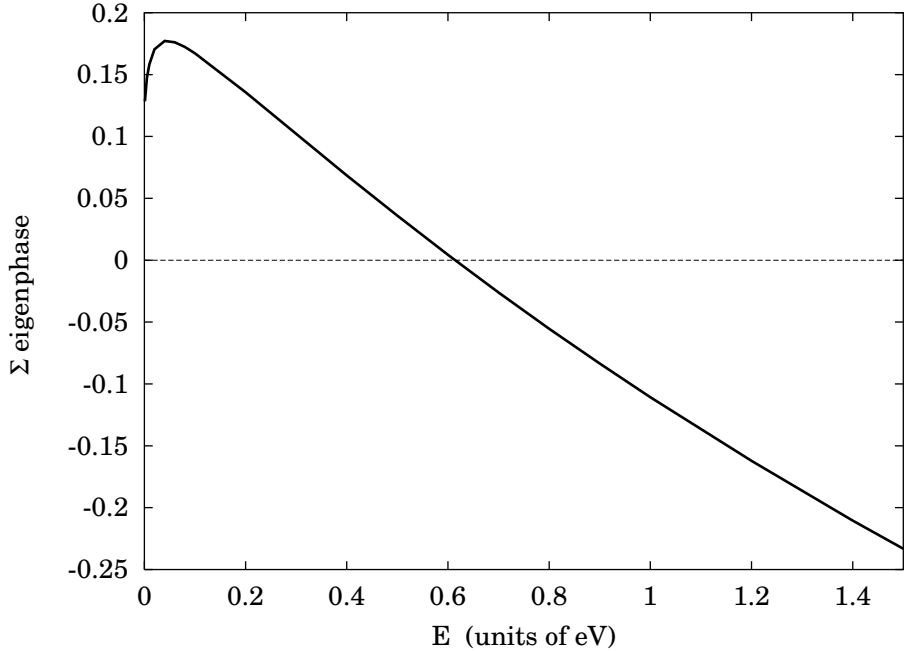}
\caption{Computed eigenphase sum behavior within the $\Sigma$
symmetry component of the scattered electron. }\label{figure2}
\end{center}
\end{figure}

\end{enumerate}

In other words, we see the dominant presence of a forward scattering
behavior of the DCSs unless we are at the resonance position, where
the 'orbiting' features of the trapped electron distorts the effects
coming from the permanent dipole of the target molecule.

\begin{figure}
\begin{center}
\includegraphics[scale=0.70]{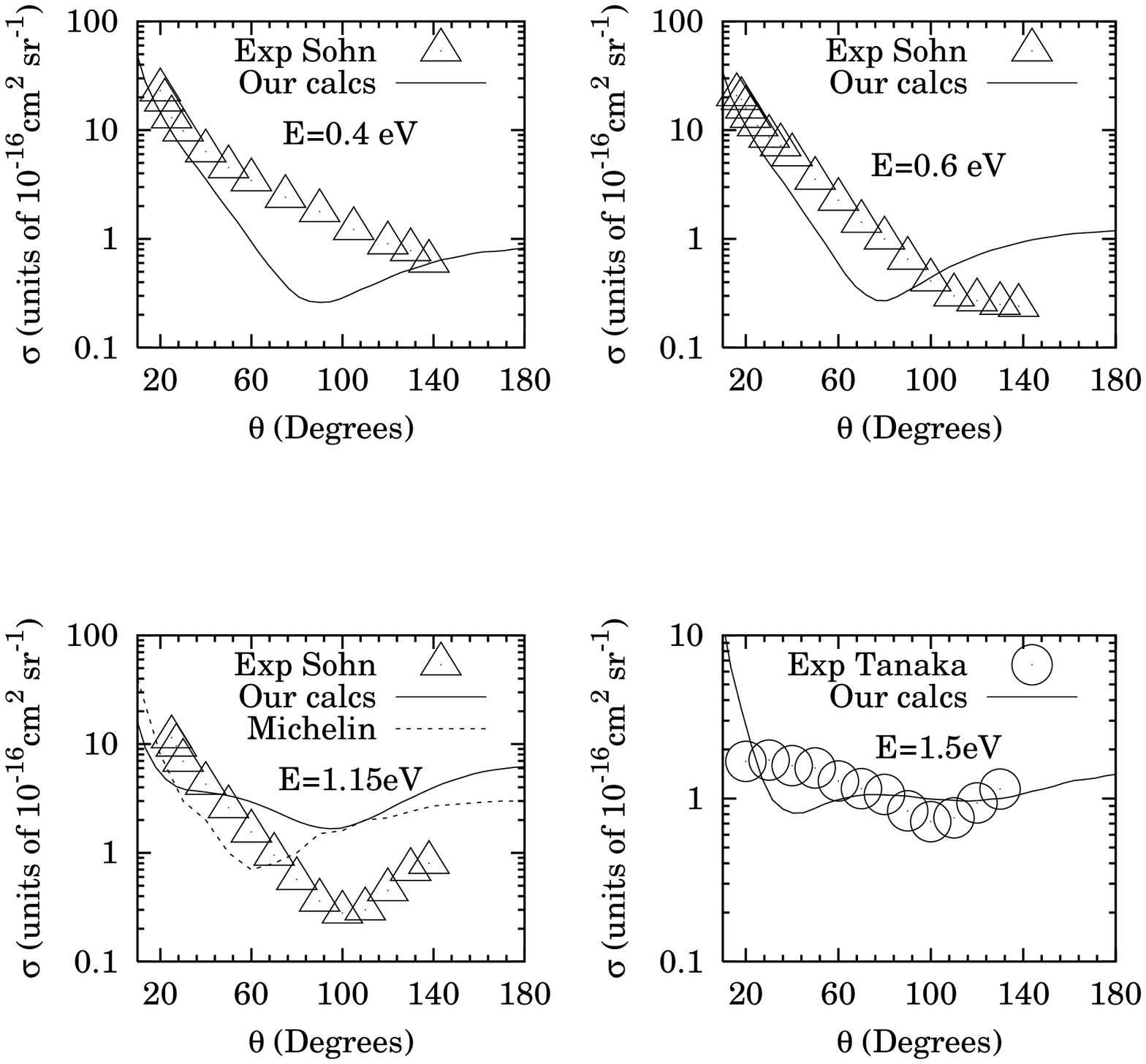}
\caption{Computed and measured angular distributions. Solid line:
present calculations, open triangles: expt.s from ref.\cite{3}. Open
circles: expt.s from ref.\cite{5}, dashes: calculations from
ref.\cite{8}. Upper left panel: 0.4 eV. Upper right panel: 0.6 eV.
Lower left panel: 1.15 eV. Lower right panel: 1.5 eV.
}\label{figure2}
\end{center}
\end{figure}

\begin{figure}
\begin{center}
\includegraphics[scale=0.70]{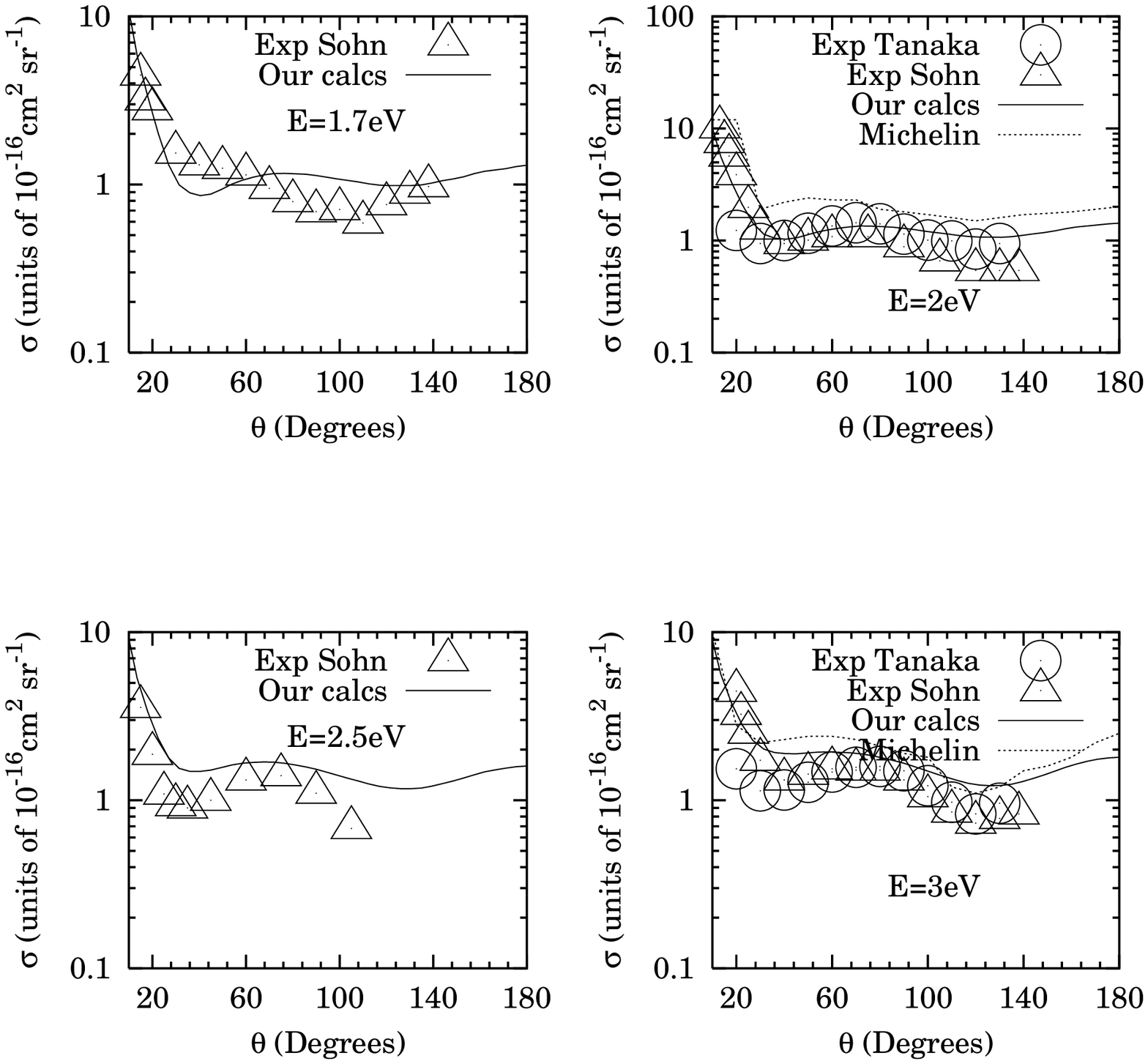}
\caption{Same as in figure 3 but for different collision energies.
The dashed lines are calculations from ref.\cite{8}. Upper left
panel: 1.7 eV; upper right panel: 2.0 eV; lower left panel: 2.5 eV;
lower right panel:3.0 eV. }\label{figure3}
\end{center}
\end{figure}

\subsection{The DCSs at higher energies}
Since the experimental data for the angular distributions are
available over a broad range of collision energies, we have analyzed
them rather carefully from above the dominant $\pi^{\ast}$ resonance
out to about 100 eV. The comparisons with the measured data are
shown in figures from 5 through 7. The following comments could be
made from a perusal of the data shown in those figures:
\begin{enumerate}
  \item the present calculations are seen to follow measurements
  remarkably well, both in shape and size, over the whole energy
  range;
  \item at 5.0 eV the data from Sohn et al.\cite{2} differ from
  those of Tanaka et al. \cite{5} in the small-angle region since
  they indicate there a strong forward peaking of the angular
  distributions: our calculated values follow those measurements
  very accurately, thereby confirming the experimental findings in
  that angular regime;
  \item as the collision energy increases one sees an increasing
  flattening of the DCSs as a function of the scattering angle and
  the appearance of the forward peaks at increasingly smaller
  angles: the calculations follow suit in the sense that indeed
  show the same general behavior as that indicated by the experiments.
\end{enumerate}

\begin{figure}
\begin{center}
\includegraphics[scale=0.70]{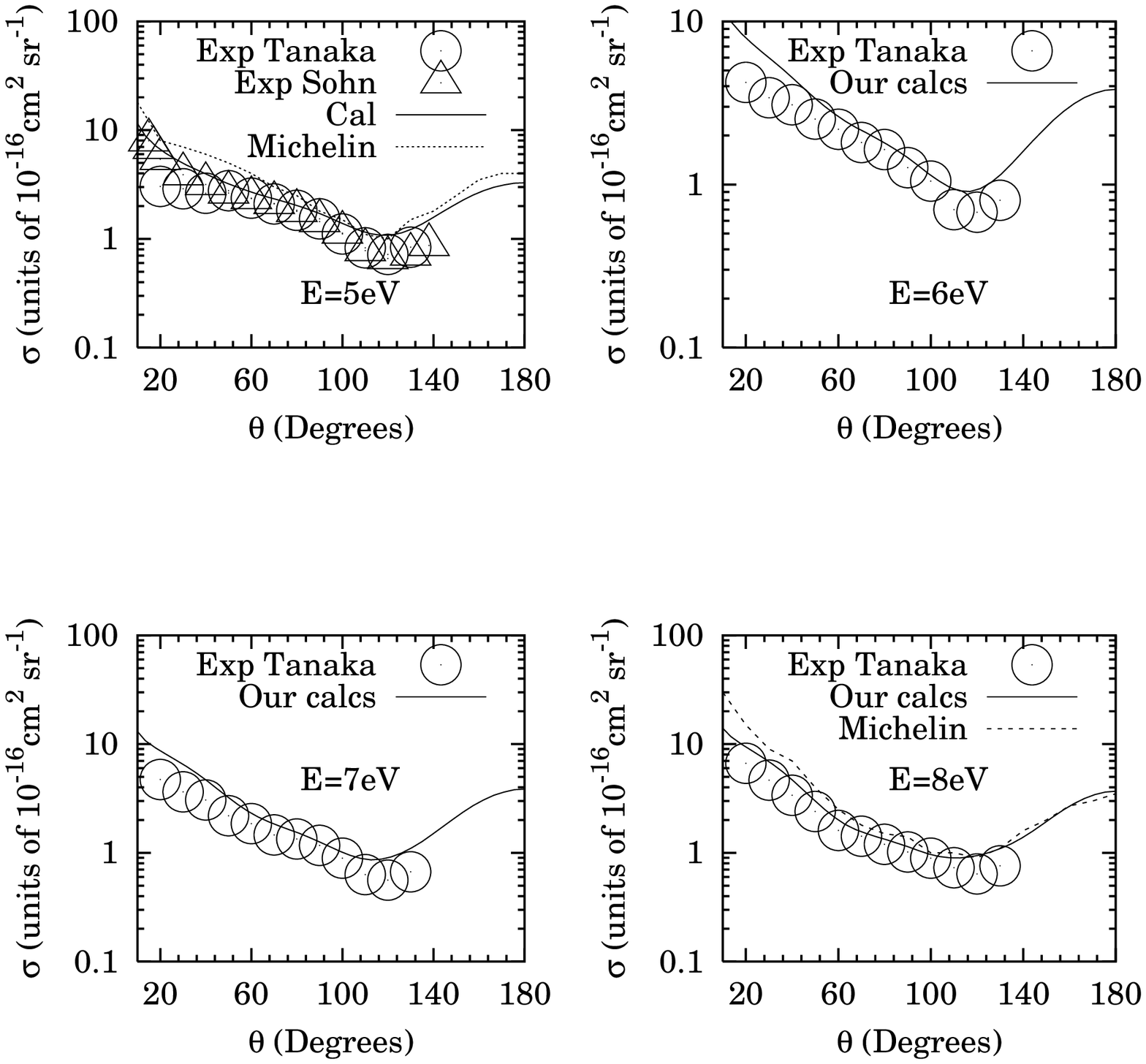}
\caption{Computed and measured angular distributions. Solid lines:
present calculations. Dashes: calculations from ref.\cite{8}. Open
circles: expt.s from ref.\cite{5}; triangles: expt.s from \cite{3}.
Upper left panle: 5.0 eV. Upper right panel: 6.0 eV. Lower left
panel: 7.0 eV. Lower right panel: 8.0 eV. }\label{figure4}
\end{center}
\end{figure}

\begin{figure}
\begin{center}
\includegraphics[scale=0.70]{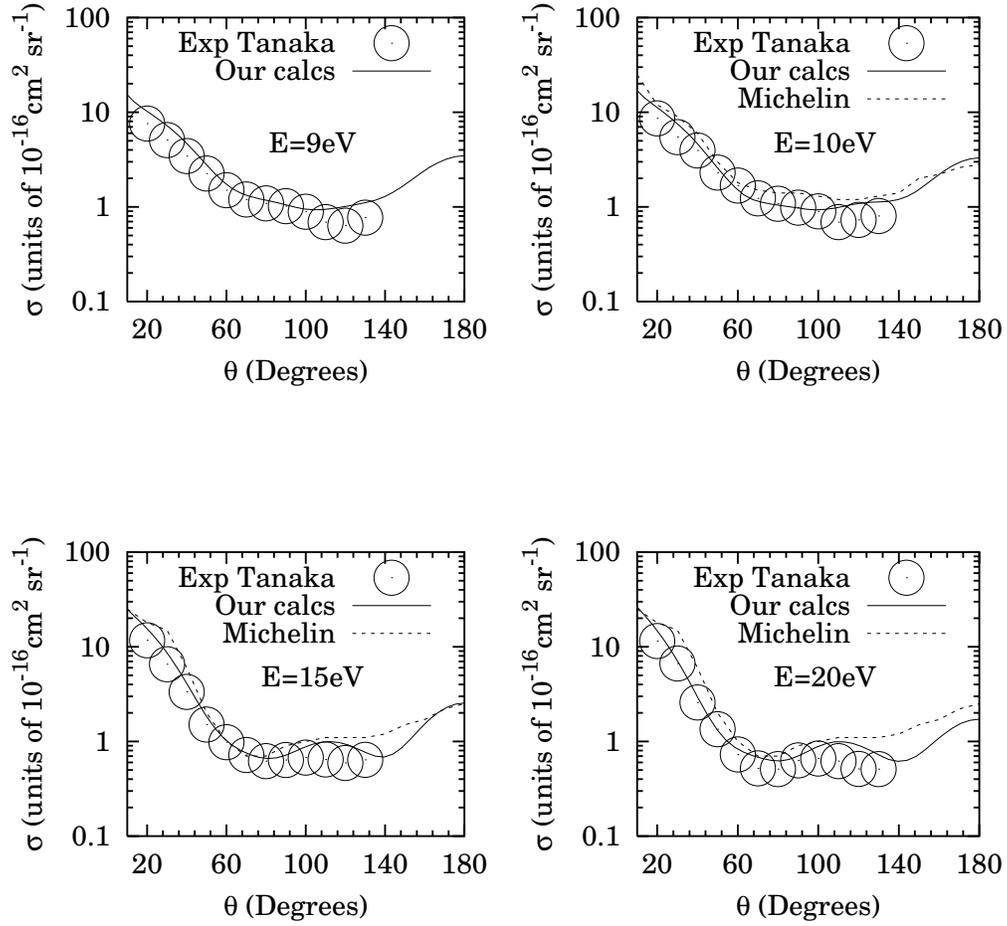}
\caption{Same as in figure 5 but for different collision energies.
Upper left panel: 9.0 eV; upper right panel: 10.0 eV; lower left
panel: 15 eV; lower right panel:20.0 eV. }\label{figure5}
\end{center}
\end{figure}

\begin{figure}
\begin{center}
\includegraphics[scale=0.50]{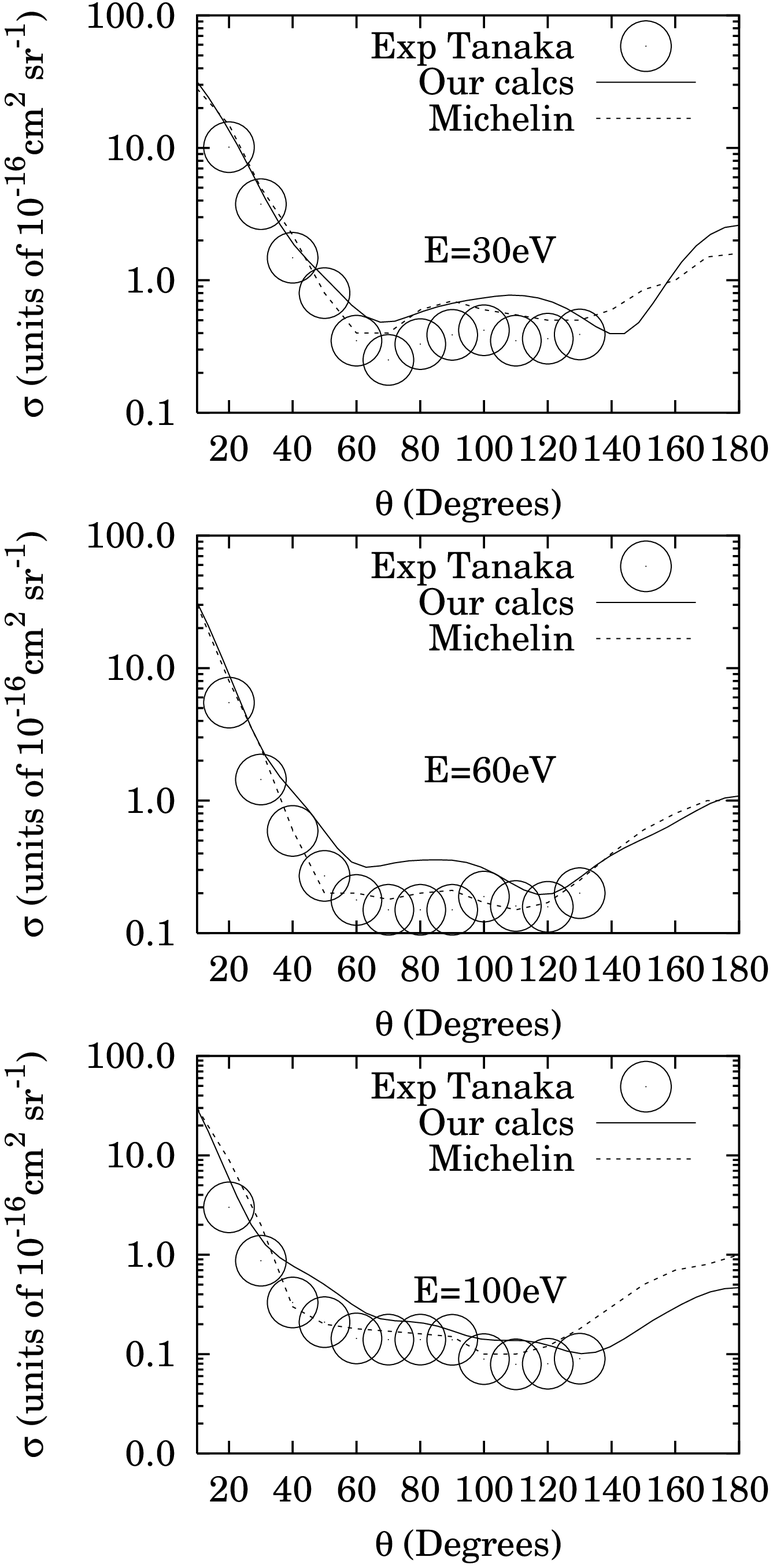}
\caption{Same as in figure 6 but for different collision energies,
Upper left panel: 30.0 eV.  Middle panel: 60.0 eV. Lower panel:
100.0 eV. }\label{figure6}
\end{center}
\end{figure}

\section{Present conclusions}
The calculations discussed in this work have analyzed, using a
quantum treatment of the scattering dynamics, the elastic cross
sections for electron impact on the OCS target molecule in the gas
phase. In particular, we have employed the exact description of the
static potential and a separable representation of the exchange
potential, both represented via a single-center expansion,  and we
have further added correlation-polarization effects via a density
functional formulation that we have discussed many times before
\cite{17}. The corresponding integro-differential equations have
been solved via quadratures of Volterra equations as discussed in
section II and the angular distributions have been computed
including the Born dipole corrections beyond the partialwave value
of  $l_{max}$=30 \cite{20}. The final results for the integral cross
sections indeed confirm the presence, around 1 eV of collision
energy, of a narrow and intense shape resonance of $\Pi$-symmetry
associated to a well-known $\pi^{\ast}$ resonance for the title
system (e.g- see discussion in ref.s\cite{7,9}. It also suggests the
presence of further two resonances of $\pi^{\ast}$ and
$\sigma^{\ast}$ symmetry and larger widths, together with a clear
Ramsauer-Townsend minimum in the cross sections around 0.7 eV
\cite{10}. Furthermore, the various available angular distributions
for the elastic scattering have also been analyzed over a very broad
range of collision energies, spanning nearly 100 eV, and compared
with the existing experiments and with the earlier calculations
\cite{8,9,10}. The comparison of all the distribution data were
presented in the previous Section and they show fair agreement
between the measurements and with other computations \cite{8,9,10}.
The present calculations also show particularly good correspondence
between computed and measured distributions for energies from 3.0 eV
and up to 100 eV. Considering the complex, many-electron structure
of the target and the broad range of collision dynamics which has
been analyzed, our results do indicate the robustness of the adopted
dynamical integrator and the reliability of the theoretical
modelling which we have employed in this study.

\section{acknowledgements}
We thank the University of Rome, the Caspur Supercomputing
Consortium and the COST Project "EIPAM" for financial and
computational support during the carrying out of the present work.

\end{document}